\documentclass[epj]{webofc}
\usepackage[varg]{txfonts}   
\wocname{EPJ Web of Conferences}
\woctitle{CONF12}
\begin{document}
\selectlanguage{english}
\title{Progress in three-particle scattering from LQCD}
%
%

\author{Ra\'ul A. Brice\~no\inst{1} \and
        Maxwell T. Hansen\inst{2}\fnsep\thanks{Speaker,\ \email{hansen@kph.uni-mainz.de}} \and
        Stephen R. Sharpe\inst{3}
}

\institute{Thomas Jefferson National Accelerator Facility, 12000 Jefferson Avenue, Newport News, VA, 23606, USA
\and
           Institut f\"ur Kernphysik and Helmholz Institute Mainz, University of Mainz, 55099 Mainz, Germany
\and
	           Physics Department, University of Washington, Seattle, WA, 98195, USA
}

\abstract{%
  We present the status of our formalism for extracting three-particle scattering observables from lattice QCD (LQCD). The method relies on relating the discrete finite-volume spectrum of a quantum field theory with its scattering amplitudes. As the finite-volume spectrum can be directly determined in LQCD, this provides a method for determining scattering observables, and associated resonance properties, from the underlying theory. In a pair of papers published over the last two years, two of us have extended this approach to apply to relativistic three-particle scattering states. In this talk we summarize recent progress in checking and further extending this result. We describe an extension of the formalism to include systems in which two-to-three transitions can occur. We then present a check of the previously published formalism, in which we reproduce the known finite-volume energy shift of a three-particle bound state.
}
\maketitle
\section{Introduction}
\label{intro}

Over the last few decades, outstanding progress has been made in extracting scattering amplitudes from the fundamental quantum theory of the strong force, QCD. This is due largely to the highly successful program of determining these quantities from numerical calculations of QCD correlators, defined on a finite discretized Euclidean space-time, using the method of lattice QCD (LQCD). 

It is not possible to determine scattering observables directly from finite-volume Euclidean correlators, and thus an indirect approach is required. The most promising technique, based on seminal work by L\"uscher, is to use the correlators to instead extract the discrete finite-volume spectrum of the theory. This can then be related to scattering amplitudes using quantization conditions of the form
\begin{equation}
\label{eq:genQC}
\Delta^\mathcal M(E, \vec P, L) = 0 \,.
\end{equation}
For fixed values of total-momentum, $\vec P$, in the finite-volume frame, and for a given linear extent of the periodic cubic volume\footnote{In this work we restrict attention to cubic, periodic finite volumes with infinite Euclidean time extent.}, $L$, this condition has a discrete list of solutions in $E$: $E_1, E_2, E_2, \cdots$, corresponding to the finite-volume energies. The superscript, $\mathcal M$, indicates that this condition depends on the on-shell scattering amplitudes of the theory. Thus for each triplet, $E_n, \vec P, L$, known to satisfy the condition, Eq.~(\ref{eq:genQC}) gives a constraint on the physical observables.

In the original work of L\"uscher \cite{Luscher1986,Luscher1990}, the quantization condition was derived for a system with a single two-particle channel of identical scalar particles. This result also assumed vanishing total momentum in the finite-volume frame and the extension to general $\vec P$ was worked out in Refs.~\cite{Rummukainen1995,Christ2005,KSS2005}. The result can be expressed as
\begin{equation}
\label{eq:origQC}
\Delta^\mathcal M(E, \vec P, L) = \mathrm{det}[\mathcal M^{-1}(E^*) + F(E, \vec P, L)] + \mathcal O(e^{- m L}) = 0 \,,
\end{equation}
where $E^* = \sqrt{E^2 - \vec P^2}$ is the center of mass (CM) frame energy. We note that L\"uscher's quantization condition, and indeed all quantization conditions discussed here, is valid up to neglected exponentially suppressed corrections scaling as $e^{- m L}$ where $m$ is the physical mass of the lightest particle in the system. This holds, as long as one restricts attention to energies below three-particle production threshold $E^*<3m$.\footnote{If even- and odd-particle numbers are decoupled then the restriction is $E^*<4m$.} Here both the scattering amplitude, $\mathcal M$, and the geometric function, $F$, are matrices with index space $\ell m$, where $\ell = 0,1,\cdots$ and $m = - \ell, - \ell+1, \cdots, \ell$. The quantity $F$ is defined in Ref.~\cite{KSS2005}. For our purposes it is sufficient to note that it is a known geometric function, so that each solution to Eq.~(\ref{eq:origQC}) gives a constraint on $\mathcal M$.


More recently the quantization condition has been extended to describe non-identical and non-degenerate particles, particles with intrinsic spin as well as any number of strongly coupled two-particle channels \cite{Bernard2008,Lage2009,Bernard2010,Doring2011,Fu2012,Gockeler2012,HSmultiLL,BricenoTwoPart2012,BricenoSpin}. Here we only give further details on the coupled-channel result. In particular, if we suppose a system with two different two-scalar channels, denoted $A$ and $B$, then the quantization condition takes the form \cite{HSmultiLL,BricenoTwoPart2012}
\begin{equation}
\label{eq:coupQC}
\Delta^\mathcal M(E, \vec P, L) = \mathrm{det} \left [ \begin{pmatrix}  \mathcal M_{AA}(E^*) & \mathcal M_{AB}(E^*) \\ \mathcal M_{BA}(E^*) & \mathcal M_{BB}(E^*)  \end{pmatrix}^{-1} + \begin{pmatrix} F_{A}(E, \vec P, L) & 0 \\  0 &  F_{B}(E, \vec P, L)   \end{pmatrix} \right ] + \mathcal O(e^{- m L}) = 0 \,.
\end{equation}
Here the individual entries are again matrices with indices $\ell m$ and the geometric functions $F_A$ and $F_B$ depend on the masses of the particles and on whether or not they are identical, but not on the dynamics of the theory. This coupled-channel formalism has been implemented numerically with great success by the Hadron Spectrum Collaboration to study a large variety of resonances that can decay into multiple two-particle channels.\footnote{For recent examples see Refs.~\cite{HadSpecDstar} and \cite{HadSpeca0}.}

\section{Three-particle formalism}

We now turn to the quantization condition for studying three-particle scattering. This was derived by two of us in Refs.~\cite{LtoK} and \cite{KtoM}. In these references we restrict attention to a system of identical scalar particles, with mass $m$, and with interactions that satisfy a $\mathbb Z_2$ symmetry so that the even- and odd-particle-number sectors are decoupled. In addition we require that the two-particle K-matrix, appearing when two of the three particles scatter in a two-to-two subprocess, be a smooth function of two-particle energy in the available kinematic range. This second limitation arose in the course of the derivation. Poles in the K-matrix generate additional finite-volume effects that we do not incorporate.

In Ref.~\cite{LtoK} we found that, in a periodic finite volume, the three-particle energies satisfy the following quantization condition:
\begin{equation}
\label{eq:QCthree}
\Delta^\mathcal M(E, \vec P, L) =  \mathrm{det}[\mathcal K_{\rm df,3}(E^*)^{-1} + F_3(E, \vec P, L)] + \mathcal O(e^{- m L}) = 0 \,.
\end{equation}
This result holds for $m < E^* < 5 m$. Here $\mathcal K_{\rm df, 3}(E^*)$ is a modified, non-standard three-particle K-matrix. We describe it in more detail below, but the most important features are that it is a hermitian matrix and is independent of $L$. By contrast, $F_3$ depends on the box size and also on the two-to-two scattering amplitude that appears as a subprocess in three-to-three scattering. Both of these matrices are defined on an index space composed of the discretized finite-volume momentum of one of the particles, $\vec k = 2 \pi \vec n/L$ with $\vec n \in \mathbb Z^3$, and the relative angular momentum of the other two, denoted $\ell m$. Thus the full index space is $k_x,k_y,k_z,\ell,m$, abbreviated $k,\ell,m$. We often refer to the momentum $\vec k$ as the spectator momentum, and the particle carrying it as the spectator.

\begin{figure}
\begin{center}
\includegraphics[scale=0.8]{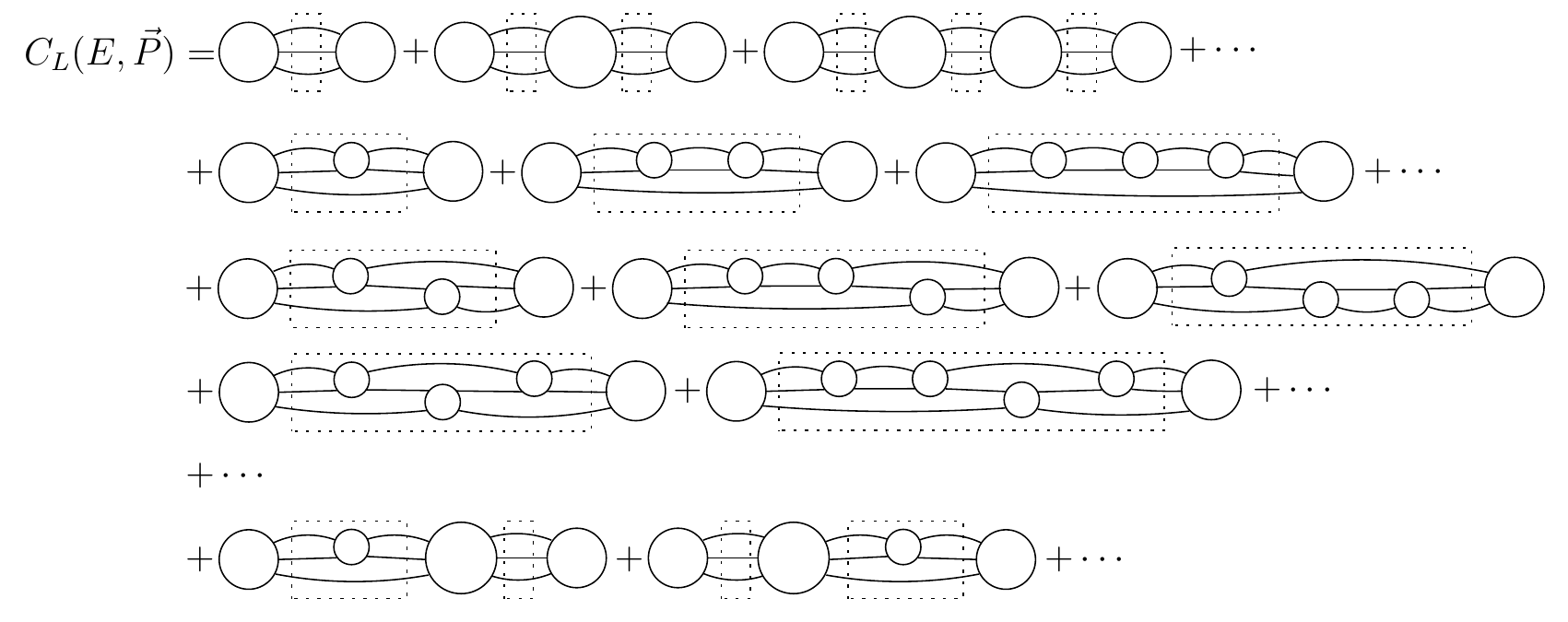}
\caption{The skeleton expansion used to display all of the $1/L$ finite-volume effects in the finite-volume correlator, $C_L(E, \vec P)$. The left-most and right-most circles in each diagram represent functions associated with the three-particle creation and annihilation operators used to define the correlator. All internal circles with four legs represent two-to-two Bethe-Salpeter kernels, and those with six legs represent a three-to-three Bethe Salpeter-like kernel, defined in Ref.~\cite{LtoK}. All of the finite-volume effects within the kernels are exponentially suppressed and, since we neglect such terms, we use the infinite-volume definitions. Thus we only incorporate the finite-volume effects from the three-particle states shown explicitly. The lines in all such states represent fully dressed propagators. Finite-volume effects arise because all loops are summed, rather than integrated, over spatial momenta $\vec k = 2 \pi \vec n/L$ with $\vec n \in \mathbb Z^3$. The dashed rectangles indicate these finite-volume sums.\label{fig:skel}}
\end{center}
\end{figure}

The three-particle quantization condition, like those for two particles, discussed in the previous section, was derived by studying a finite-volume correlator, $C_L(E, \vec P)$, using a skeleton expansion to identify all finite-volume effects that vanish more slowly than $e^{- m L}$ [see Fig.~\ref{fig:skel}]. It turns out that such important finite-volume contributions arise from sums over loops involving three particles that together carry the total energy and momentum $(E, \vec P)$. To derive Eq.~(\ref{eq:QCthree}) we write each diagram as its infinite-volume counterpart plus a finite-volume residue. Summing this to all orders gives an expression for the difference between the finite- and infinite-volume correlators, $C_L(E, \vec P) - C_{\infty}(E, \vec P)$. This difference scales as the inverse of the matrix appearing inside the determinant in Eq.~(\ref{eq:QCthree}), implying that $C_L(E, \vec P)$ has a pole, and thus $E$ is in the finite-volume spectrum, whenever the quantization condition is satisfied.

The precise definition of $F_3$ is
\begin{align}
F_3 & \equiv \frac{F}{6 \omega L^3} - \frac{F}{2 \omega L^3} \frac{1}{1 + \mathcal M_{2,L} G} \mathcal M_{2,L} F \,, \\
\mathcal M_{2,L} & \equiv \frac{1}{1 + \mathcal M_2 F} \mathcal M_2 \,,
\end{align}
where $\omega$, $F$, $G$, $\mathcal M_{2,L}$ and $\mathcal M_2$ are all matrices on the $k,\ell,m$ index space.\footnote{It turns out that the matrices $F$ and $\omega^{-1}=\frac{1}{\omega}$ commute so that we can write $\frac{F}{\omega}$ with no ambiguity.} For example, $\mathcal M_2$ is the standard infinite-volume two-to-two scattering amplitude, defined on the index space as
\begin{equation}
\mathcal M_{2; k' \ell' m'; k \ell m} = \delta_{k', k} \mathcal M_{2; \ell' m'; \ell m}(E_{2,k}^*) \,,
\end{equation}
where
\begin{equation}
E_{2,k}^* = \sqrt{(E - \omega_k)^2 - (\vec P - \vec k)^2} \,,
\end{equation}
with $\omega_k = \sqrt{\vec k^2 + m^2}$. 

\begin{figure}
\begin{center}
\includegraphics[scale=0.4]{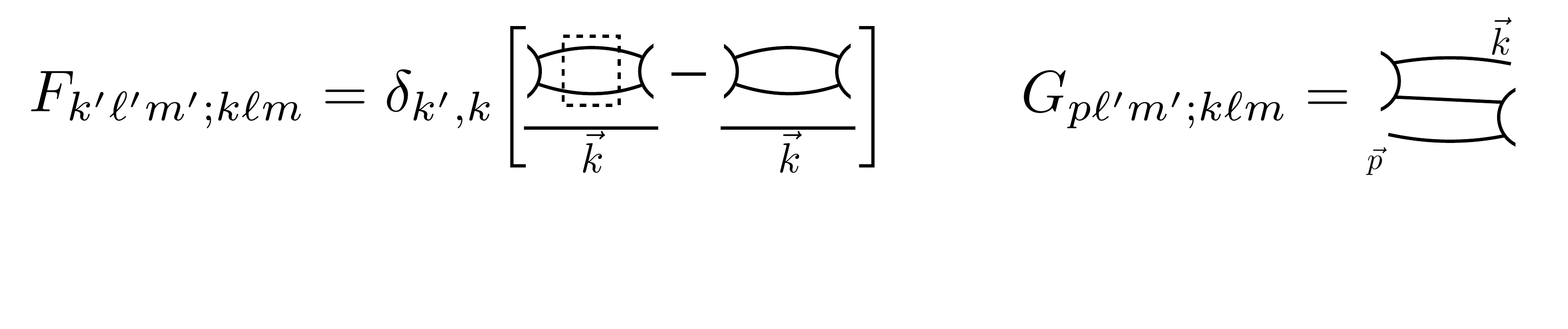}
\vspace{-40pt}
\caption{Diagrammatic representation of the two finite-volume kinematic functions, $F$ and $G$. $F$ is defined as the difference between the sum and integral over a two-particle loop where the third particle spectates with a fixed momentum $\vec k$. As is discussed in the text, the integral is defined with an analytically continued principle-value pole prescription and with a smooth cutoff function, $H(\vec k)$. $G$ is the factor that arises when a two-to-two insertion changes to a different particle pair. This matrix also has off-diagonal elements in spectator momentum, as the momenta can differ between the incoming and outgoing spectator particles.  \label{fig:FGfig}}
\end{center}
\end{figure}

The quantities $F$ and $G$ encode the finite-volume effects that arise from the sums over loops [see Fig.~\ref{fig:FGfig}]. $F$ is similar to the quantity bearing the same name introduced in the two-particle quantization conditions, Eqs.~(\ref{eq:origQC}) and (\ref{eq:coupQC}). To explain it in more detail we first give an alternative form of the original two-particle quantization condition, Eq.~(\ref{eq:origQC}),
\begin{equation}
\Delta^\mathcal M(E, \vec P, L) = \det \left [ \mathcal K_2(E^*)^{-1} + F^{\mathrm{PV}}(E, \vec P, L) \right ] + \mathcal O(e^{- m L}) = 0 \,,
\end{equation}
where $F^{\mathrm{PV}} = \mathrm{Re} F$ for $E^*> 2 m$ and for $E^* < 2 m$ we define $F^{\mathrm{PV}}$ by analytic continuation. This turns out to be equivalent to defining an integral appearing in the definition of $F$ with a principle-value pole prescription. An analogous definition holds for $\mathcal K_2(E^*)^{-1}$: $\mathcal K_2(E^*)^{-1} = \mathrm{Re} \mathcal M_2(E^*)^{-1}$ for $E^*> 2 m$ with the subthreshold extension given again by analytic continuation. This corresponds to the standard definition of the K-matrix. We stress that we have simply canceled the imaginary parts between the two terms of Eq.~(\ref{eq:origQC}) in order to reach this alternative form. It is not obvious, but one can prove the imaginary parts are equal in magnitude with opposite sign, for all values of $E, \vec P, L$.

The quantity $F$ that appears in the three-particle quantization condition is then given by
\begin{equation}
F_{k' \ell' m'; k \ell m}(E, \vec P, L) \equiv \delta_{k',k} H(\vec k) F^{\mathrm{PV}}_{\ell' m'; \ell m}(E - \omega_k, \vec P - \vec k, L) \,.
\end{equation}
We identify four key differences between the $F$ factors in the two- and three-particle quantization conditions: First the additional index dependence is included by simply multiplying on a Kronecker delta in the momentum index. Second, we use the PV definition in which the imaginary part has been cancelled above threshold, with analytic continuation below. Third, the quantity is evaluated at energy and momentum given by the total $(E, \vec P)$ minus the energy-momentum of the spectator particle $(\omega_k, \vec k)$. Fourth, and finally, we have included a new function $H(\vec k)$. This is a smooth cutoff function defined to equal $1$ for $(2m)^2 < E^{*2}_{2,k}$, $0$ for $E^{*2}_{2,k} < 0$ and to smoothly interpolate between. Including this cutoff function is necessary to avoid dependence on the two-to-two scattering amplitude far below threshold. We denote the combined effect of the analytically continued principle-value prescription and the cutoff function by $\widetilde {\rm PV}$.

This now leaves $\omega$ and $G$ for which we do not give the detailed definitions but instead only the basic idea. The matrix $\omega$ is diagonal and populated by the kinematic factor $\omega_k$ evaluated at momenta indicated by the spectator momentum index. $G$ is off diagonal in both spectator and angular momenta and encodes the finite-volume effects of a two-to-two insertion switching to a different scattering pair [see again Fig.~\ref{fig:FGfig}]. The full definitions of all quantities are given in Ref.~\cite{LtoK}.

If the two-particle quantization condition is used to determine $\mathcal M_2$ for two-particle CM energies below $4m$, then all of the building blocks entering $F_3$ are known and the quantity can be determined. Practically this requires truncating the matrices in angular momentum space, which is expected to be a good approximation at low energies. Then, given $F_3$, each energy in the three-particle spectrum gives a constraint on $\mathcal K_{\rm df, 3}$. Given a particular model or parametrization of the latter, and a large number of energy levels, one can then in principle extract this infinite-volume quantity. However, $\mathcal K_{\rm df, 3}(E^*)$ is a non-standard K-matrix. This is the central shortcoming of the quantization condition, Eq.~(\ref{eq:QCthree}), derived in Ref.~\cite{LtoK}. 

This issue was resolved in Ref.~\cite{KtoM} where we derived an integral equation relating $\mathcal K_{\rm df,3}$ to the standard, relativistic three-to-three scattering amplitude, denoted $\mathcal M_3$. The equation was derived by studying an alternative finite-volume correlator, denoted $\mathcal M_{3,L}$. This new correlator has the same pole structure as $C_L$ and depends on $\mathcal K_{\rm df, 3}$ and $F_3$ in a similar way. But $\mathcal M_{3,L}$ is also defined to become the standard infinite-volume scattering amplitude in a carefully constructed infinite-volume limit. 

Expressing $\mathcal M_{3,L}$ in terms of $\mathcal K_{\rm df, 3}$ and taking the limit, $L \to \infty$, one recovers an integral equation relating the non-standard K-matrix to the physically observable scattering amplitude. The result is
\begin{equation}
\mathcal M_3 = \lim_{\epsilon \to 0} \lim_{L \to \infty} \mathcal S \left[\mathcal D_L^{(u,u)} + \mathcal L_L^{(u)} \mathcal K_{\rm df,3}  \frac{1}{1 + F_3 \mathcal K_{\rm df,3}}   \mathcal R_L^{(u)}    \right ] \,.
\end{equation}
Here the series of limits indicates that one must evaluate the energy $E$ in all poles appearing in sums at a shifted value $E \to E + i \epsilon$, then send $L \to \infty$ with $\epsilon$ fixed and positive, and then send $\epsilon \to 0$. The quantities $\mathcal D_L^{(u,u)}$, $\mathcal L_L^{(u)}$ and $\mathcal R_L^{(u)}$ appearing here are closely related to $F_3$, with the exact definitions given in Ref.~\cite{KtoM}. Finally, the $u$ superscripts indicate that these quantities are unsymmetrized, meaning that they are not invariant under interchange of external momenta. The operator $\mathcal S$ stands for a symmetrization procedure that gives the correct result of $\mathcal M_3$.

This completes the summary of our three-particle quantization condition. The two major limitations of the result, mentioned at the beginning of this section, are the assumed $\mathbb Z_2$ symmetry that decouples even- and odd-particle-number sectors, and the assumed smoothness of the two-particle K-matrix appearing in two-to-two subprocesses \cite{noZ2}. In the following section we describe the removal of this first assumption. Incorporating two-particle K-matrix poles is also under investigation, but we do not discuss this work further in the present write-up.

\section{Extension to include two-to-three transitions}
\label{sec:twotothree}

In this section we describe the extension of the three-particle quantization condition discussed above, to include systems with arbitrary $\textbf 2 \to \textbf 3$ transitions. The set-up is otherwise the same as in the previous section. In particular we assume identical scalar particles with mass $m$ in a periodic volume of extent $L$. We also continue to assume no poles in the two-particle K-matrices. One important distinction is that, with the $\mathbb Z_2$ symmetry removed, we must change the kinematic region considered. In particular we restrict attention to $m < E^* < 4m$, the range in which, at most, two- and three-particle states can go on shell.

\begin{figure}
\begin{center}
\includegraphics[scale=0.5]{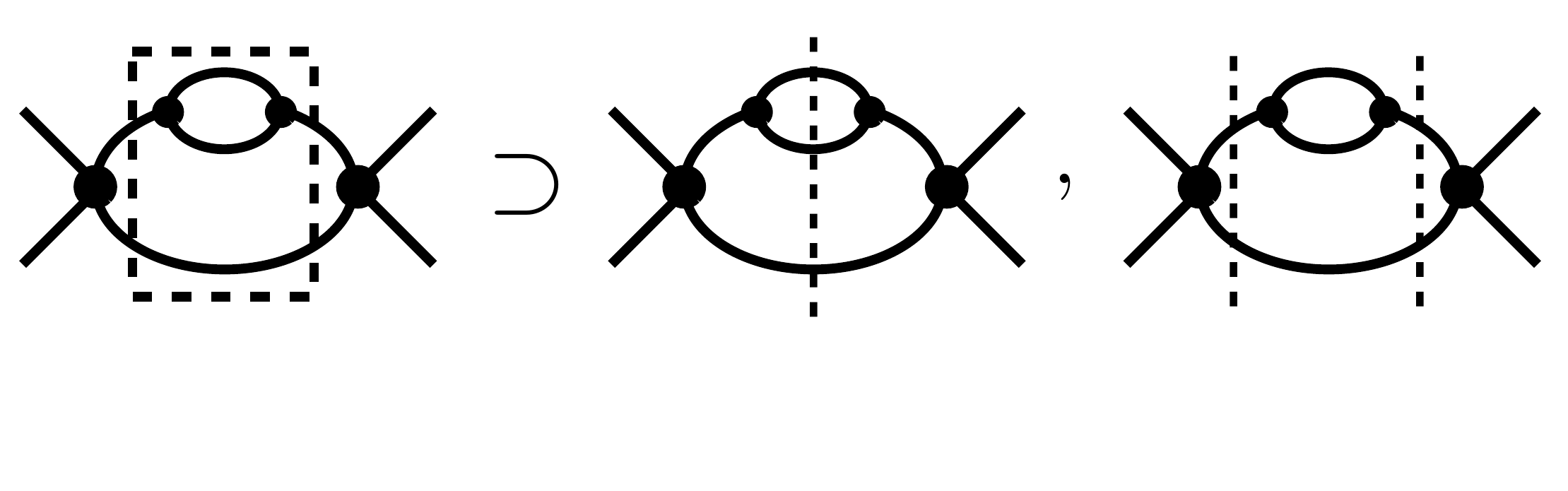}
\vspace{-30pt}
\caption{Example of a finite-volume Feynman diagram that gives rise to both two- and three-particle states. These types of diagrams are a particular source of complication because (1) they include on-shell states within self-energy diagrams, implying that certain self-energy contributions must be explicitly displayed and (2) they involve disconnected two-to-three transitions mediated by a one-to-two coupling with a spectator.  \label{fig:cuts}}
\end{center}
\end{figure}

The skeleton expansion shown in Fig.~\ref{fig:skel} cannot be used in the present analysis since it omits all of the diagrams in which two particles carry the total energy and momentum. Indeed significant new complications arise when one attempts to extend the skeleton expansion to the system with $\textbf 2 \to \textbf 3$ transitions. This is due primarily to diagrams such as that shown in Fig.~\ref{fig:cuts}. Here the right-hand side represents a particular finite-volume diagram contributing to the correlator. The vertical dashed lines on the right-hand side show that the diagram contains both two- and three-particle states that lead to important finite-volume effects.

As a result of these types of diagrams, it is no longer possible to use fully dressed propagators in the diagrammatic expansion of the finite-volume correlator. The two-particle bubbles that arise on one of the propagators in a two-particle loop carrying $(E, \vec P)$ must be explicitly displayed, since these lead to on-shell three-particle states with important finite-volume effects. This implies that the two-particle loops must be expressed in terms of two-particle irreducible (2PI) propagators, defined by summing all 2PI diagrams into a 2PI self energy, and defining a propagator from this using the standard relation. 

A second issue associated with the diagram in Fig.~\ref{fig:cuts} is that the $\textbf 2 \to \textbf 3$ transition occurs via a $\textbf 1 \to \textbf 2$ subprocess with a spectator particle. These disconnected transitions are complicated because they result in a common coordinate, the spectator, appearing in both the two- and three-particle states, making it more difficult to separately identify the finite-volume contributions. However, such disconnected transitions do not appear in our final result, due to the fact that they are kinematically forbidden for on-shell states. Showing how these terms fall out of the final result is nontrivial and is another key result of Ref.~\cite{noZ2}.

In Ref.~\cite{noZ2} we work out the full details of a diagrammatic expansion that allows us to explicitly display all finite-volume effects for the on-shell two- and three-particle states. We find it convenient to depart from the skeleton expansion approach of Refs.~\cite{LtoK} and \cite{KtoM} and instead work with diagrams expressed in terms of contact interactions and both fully-dressed and 2PI-propagators, depending on the diagram topology.\footnote{In fact 3PI propagators must also be incorporated in the case where a single propagator carries the entire energy-momentum $(E, \vec P)$. However these generate no important finite-volume effects and thus play only a minor role in the derivation.} We then study all possible diagrams in time-ordered perturbation theory and determine the form of all two- and three-particle poles leading to $1/L^n$ finite-volume effects.

This analysis allows us to piggyback on previous results in the two- \cite{Luscher1986,Luscher1990,Rummukainen1995,Christ2005,KSS2005} and three-particle \cite{LtoK,KtoM} sector. However, various technical issues arise that require careful examination as we explain in Ref.~\cite{noZ2}. In the end, the modified quantization condition for this new system takes an intuitive form
\begin{equation}
\Delta^\mathcal M(E, \vec P, L) = \mathrm{det} \left [ \begin{pmatrix}  \mathcal K_{22}(E^*) & \mathcal K_{23}(E^*) \\ \mathcal K_{32}(E^*) & \mathcal K_{\rm df,33}(E^*)  \end{pmatrix}^{-1} + \begin{pmatrix} F_{2}(E, \vec P, L) & 0 \\  0 &  F_{3}(E, \vec P, L)   \end{pmatrix} \right ] + \mathcal O(e^{- m L}) = 0 \,.
\end{equation}
This resembles the coupled two-particle quantization condition, Eq.~(\ref{eq:coupQC}), with the three-particle result in the lower diagonal entries.

As with the three-particle quantization condition in the case of a $\mathbb Z_2$ symmetry, here the two-by-two K-matrix is a non-standard object in all of its entries. One therefore again requires integral equations to convert it to the standard scattering amplitude in the coupled two- and three-particle sectors. We present the derivation of these in Ref.~\cite{noZ2} as well, and thus give the complete relation between finite-volume energies and scattering amplitudes in the combined two- and three-particle system.

\section{Three-particle bound state}
\label{sec:bound}

In addition to extending the results of Refs.~\cite{LtoK} and \cite{KtoM}, it is also important to provide nontrivial checks. This is necessary because the derivation is complicated and because such checks also shed light on the utility of the quantization condition. Recently, two of us performed two such checks, studying the quantization condition in the limit of very weak \cite{weak} and very strong \cite{strong} interactions. In both cases we were able to reproduce and extend existing results derived in non-relativistic quantum mechanics.

In this section we summarize the work of Ref.~\cite{strong}, in which we reproduce the finite-volume energy shift to a spin-zero three-particle bound state in a system with two-particle interactions at unitarity (divergent scattering length). This system was studied by Mei{\ss}ner, R\'ios and Rusetsky (MRR) using non-relativistic quantum mechanics. Defining the binding momentum $\kappa$ via $E_B = 3 m - \kappa^2/m$, where $E_B$ is the infinite-volume bound state energy (mass), MRR found \cite{MRR}
\begin{equation}
\label{eq:MRRresult}
\Delta E_B(L) \equiv E_B(L) - E_B = c \vert A \vert^2 \frac{\kappa^2}{m} \frac{1}{(\kappa L)^{3/2}} e^{- 2 \kappa L/\sqrt{3}}   + \cdots \,,
\end{equation}
where $E_B(L)$ is the energy of the finite-volume state (in a periodic cubic volume with extent $L$) that corresponds to the bound state in the infinite-volume limit. The ellipses indicates terms that vanish more quickly as $L \to \infty$. Here $c$ is a known geometric constant, arising from the Efimov bound-state wavefunction, and $A$ is a parameter related to the normalization of the wavefunction that is close to unity provided long-range effects dominate in the creation of the bound state \cite{MRR}.

The three-particle quantization condition of Refs.~\cite{LtoK} and \cite{KtoM} holds for any system, provided the restrictions listed above are satisfied. In particular, if we choose the two- and three-particle scattering amplitudes to correspond to a theory at unitarity with a three-particle bound state, then the lowest lying energy predicted by the quantization condition, Eq.~(\ref{eq:QCthree}), must be the level corresponding to the bound state, shifted by the finite volume. 

In Ref.~\cite{strong} we perform this exercise, substituting
\begin{equation}
\label{eq:amps}
\mathcal M_2 = - \frac{16 \pi E_2^*}{i p^*_2} \,, \ \ \ \ \ \mathcal M_3(p,k) \sim - \frac{\Gamma(p) \, \overline \Gamma(k)}{E^2 - E_B^2} \,,
\end{equation}
and then studying the lowest lying finite-volume level. Here $E_2^*$ is the CM energy of the two particles and $p_2^*$ the magnitude of momentum for one of the two. In the three-to-three scattering amplitude, $\Gamma(p)$ and $\overline \Gamma(k)$ are matrix elements coupling the scalar bound state to the three-particle asymptotic states. To reach this form we assume, following MRR, that the particle pair described by $\mathcal M_2$ as well as two of the three particles in the three-particle state described by $\mathcal M_3$ are dominated by s-wave interactions. This leaves only the dependence on the momentum of the third particle and, because the bound state is a scalar so that no direction is defined in the CM frame, $\Gamma(p)$ and $\overline \Gamma(k)$ can only depend on the magnitude of the spectator momentum, as shown.

Reproducing Eq.~(\ref{eq:MRRresult}) from our quantization condition proceeds in three basic steps. First, we show that our quantization condition implies
\begin{equation}
\label{eq:QCbound}
\Delta E_B(L) = - \frac{1}{2 E_B} \bigg [ \frac{1}{L^3} \sum_{\vec k} - \int_{\vec k}  \bigg ] \frac{\overline \Gamma^{(u)}(k) \,\Gamma^{(u)}(k) }{2 \omega_k \mathcal M_2(E^*_{2,k})}  + \cdots \,,
\end{equation}
where $\int_{\vec k} = \int d^3 k/(2 \pi)^3$ and the sum runs over all momenta in the finite-volume set. The $\Gamma$ quantities here carry a superscript to indicate that they are unsymmetrized versions of those defined in Eq.~(\ref{eq:amps}). This relation is not particularly intuitive but can be derived by studying our quantization condition in the vicinity of the three-particle bound-state pole, as we describe in detail in Ref.~\cite{strong}.

Second, we derive the functional form of the two-to-two scattering amplitude, by substituting the definition of $E^*_{2,k}$
\begin{equation}
\label{eq:M2final}
\mathcal M_2(E^*_{2,k}) = \frac{32 \pi m}{\kappa} \left[ 1 + \frac{3 k^2}{4 \kappa^2}  \right ]^{-1/2} \,,
\end{equation}
where we have dropped terms that are suppressed by higher powers of $\kappa^2/m^2$ or $k^2/m^2$. In addition we derive the functional form of $\Gamma^{(u)}(k)$ by matching to the Efimov bound-state wavefunction
\begin{equation}
\label{eq:Gamform}
\Gamma^{(u)}(k) = \frac{3^{3/8} \pi^{1/4}}{4} A \sqrt{-c} \mathcal M_2(E^*_{2,k}) \,.
\end{equation}
It turns out that the geometric constant $c$ is negative so that the argument of the square-root here is positive.

Third, and finally, we substitute Eqs.~(\ref{eq:M2final}) and (\ref{eq:Gamform}) into Eq.~(\ref{eq:QCbound}) and evaluate using the Poisson summation formula. We deduce Eq.~(\ref{eq:MRRresult}), in perfect agreement with MRR.

In addition to reproducing the known result, in Ref.~\cite{strong} we also generalize the expression to the case where the bound state carries nonzero momentum, $\vec P$, in the finite-volume frame. Given our derivation, the generalization to moving frames turns out to be a straightforward kinematic extension. As we show in Ref.~\cite{strong}, the result is
\begin{equation}
\Delta E_{B, \vec P}(L) = \sqrt{E_B(\vec P, L)^2 - \vec P^2} - E_B = f_3[\vec n_P] \Delta E_B(L) + \cdots \,,
\end{equation}
with
\begin{equation}
f_3[\vec n_P] = \frac{1}{6} \sum_{\hat s} e^{i (2 \pi/3) \hat s \cdot \vec n_P} \,,
\end{equation}
where the sum runs over the six unit vectors with integer components. Here $\vec n_P \equiv \vec P L/(2 \pi)$ and the result assumes $\vec n_P = \mathcal O(L^0)$. It is particularly interesting to note that $f_3[(0,1,1)]=0$, meaning that the leading finite-volume shift vanishes for the bound state in this particular frame.

\section{Conclusions}

In this note we have reviewed the status of extending finite-volume quantization conditions to the three-particle sector. We summarized the result of Refs.~\cite{LtoK} and \cite{KtoM}, where the quantization condition was derived for a system of identical scalar particles. This result has two key assumptions: (1) absence of $\textbf 2 \to \textbf 3$ couplings and (2) smoothness of the two-particle K-matrix. In Sec.~\ref{sec:twotothree} we described work, nearing completion, to lift the first assumption. Removal of the second assumption, leading towards a generic two- and three-particle quantization condition, is underway.

Finally, in Sec.~\ref{sec:bound}, we described a test of the formalism published in Refs.~\cite{LtoK} and \cite{KtoM}. We outlined how the result was used to reproduce, and extend, the known finite-volume shift to a three-particle bound state. The extension to bound states with nonzero momentum in the finite-volume frame shows that the leading shift vanishes for certain momenta, making these frames potentially more useful for directly extracting the infinite-volume binding energy.

Looking forward, once the formal foundation is established, the next step is to implement the formalism in a numerical calculation. Only by applying these relations using actual LQCD data, can one establish the utility of this tool for extracting three-particle scattering and resonance properties from the underlying theory.

\bibliography{ref}

\end{document}